# Roman CCS White Paper

# Enabling Stellar Flare Science in the Roman Galactic Bulge Survey: Cadence, Filters, and the Read-Out Strategy Matter

**Roman Core Community Survey:** Galactic Bulge Time Domain Survey

**Scientific Categories:** stellar physics and stellar types

**Additional scientific keywords:** low-mass stars

**Submitting Author:**
Name: Guadalupe Tovar Mendoza
Affiliation: University of Washington, Seattle
Email: tovarg@uw.edu

**List of contributing authors**:
Guadalupe Tovar Mendoza [University of Washington, Seattle] [tovarg@uw.edu],
Robert F. Wilson [NASA GSFC] [robert.f.wilson@nasa.gov],
Allison Youngblood [NASA GSFC] [allison.a.youngblood@nasa.gov],
Laura D. Vega [NASA GSFC] [laura.d.vega@nasa.gov],
Thomas Barclay [NASA GSFC] [thomas.barclay@nasa.gov],
James R. A. Davenport [University of Washington, Seattle] [jrad@uw.edu],
Jordan Ealy [University of Maryland, College Park] [jordan.ealy@nasa.gov]

**Abstract:** As was discovered with other wide field, precise imagers, the stable photometry necessary for the microlensing surveys is well-suited to general stellar astrophysics, including stellar flares, which are important for understanding stellar magnetic activity and even the space weather environments of exoplanets. Large stellar flare surveys have never been performed before in the Roman spectral range, and Roman may reveal new information about flare emission mechanisms (blackbody, recombination continuum, chromospheric emission lines) and how flare rates change with stellar age and metallicity. For instance, the Galactic Bulge stars will be much older than the typical studied flare stars, and Roman's wide field and exquisite imaging may provide sufficient statistics to probe the flare behavior and properties of such an old stellar population. However, the information yield will likely depend on sky location, cadence, read-out strategy, and filter choices. Stellar flare timescales range from seconds to hours, so Roman may only be able to resolve the longest, most energetic events. However, because a single exposure in the Galactic Bulge Time Domain Survey will consist of several non-destructive reads, short duration events can be modeled from the flux variations within a single exposure. Here we provide a proof of concept, showing that flare morphologies can be significantly better constrained if sub-exposure data are analyzed. As a result, we advocate that such data be made publicly available for Roman flare studies, with minimal on-board processing.

**Motivation and Overview:**
Stellar flares are eruptive variables that rapidly increase in brightness, on minutes to hours long timescales. They have been the subject of intensive study over the past decade due to an abundance of wide-field, long-baseline, high-precision photometry from Kepler/K2 and TESS. Stellar flares are caused by magnetic reconnection events that inject significant energy into the outer atmospheres of stars, accelerating charged particles and releasing large amounts of energy across the electromagnetic spectrum (Benz & Gudel 2010). Intense, frequent flares and accompanying energetic particles can result in atmospheric escape by ionizing and expanding a planet's upper atmosphere (e.g., Segura et al. 2010; Shields et al. 2016; Tilley et al. 2019). This can have a significant impact on the habitability of a planet, reducing the planet's ability to retain water and maintain a stable climate (Luger et al. 2015). The study of magnetic stellar activity is therefore an essential component of exoplanet research, as it allows us to better understand the conditions of planet habitability and assess the likelihood of life beyond our solar system.

The Nancy Grace Roman Space Telescope (Roman) takes the next step, observing deeper and with a spatial resolution orders of magnitude better than Kepler and TESS, allowing us to *measure flares from stellar populations well outside the solar neighborhood for the first time*. This will enable the study of magnetic stellar activity of Sun-like stars at low metallicities and beyond "field age" (~5 Gyr) because stars in the Galactic bulge and the innermost regions of the disk are older and span a much broader range of stellar metallicities than stars in the solar neighborhood (e.g., Zoccali et al. 2017; Buck et al. 2018; Eilers et al. 2022). Stellar flares are a manifestation of a star's dynamic surface magnetism, which is strongly dependent on stellar age (Davenport et al. 2019). As a result, the flare rates of old stellar populations are expected to be extremely low, but are essentially unconstrained. While reduced flare rates with increasing stellar age have been seen in the solar neighborhood (albeit with some exceptions, e.g., Proxima Centauri and Barnard's Star), there are too few stars in the solar neighborhood whose ages exceed 10 Gyr for a robust estimate. The long time baseline combined with the overwhelming number of monitored stars in the Roman Galactic Bulge Time Domain Survey (GBTDS), provides the ideal testing ground for probing the flare rates of such old stars. This is particularly true for high-energy events. Stellar flares are known to follow a decreasing power law distribution with energy $>10^{31}$ erg (e.g. Hawley et al. 2014) meaning that such events are rare, but they are detectable in long cadence data (e.g., Davenport et al. 2016; vanDoorsselaere et al. 2017). Given the sheer unprecedented sample size of the GBTDS, many such events should be confidently detected. Furthermore, these Roman flare observations will contribute to our understanding of the relationships between flaring and stellar spectral type, luminosity, age, and rotation (e.g., Davenport et al. 2019; Ilin et al. 2021).

**Potential impact of read-out strategy, filters, and cadence on Roman flare science in GBTDS:**
Stellar flare detections from foreground (disk) and bulge stars in Roman's GBTDS will be both exciting for flare star research, yet undesirable for other studies where flares may be a source of contamination. To prepare for and enhance the science return of Roman, we must consider how Roman's cadence, read-out strategy, and filter choices impact our ability to detect stellar flares.

***Cadence:*** Results from the TESS mission have highlighted the importance of shorter cadence observations for flare studies. At higher time resolution, the measured flare energies are more accurate and finer structure of flare events is revealed (e.g., Howard et al. 2022; Tovar Mendoza et al. 2022), such as the presence of multi-peak or complex flares. Even though Roman's cadence will not be able to resolve the most common, lower-energy events, it will be able to detect them. For the high-energy events with durations greater than 30 minutes (twice the observing cadence) Roman will be able to measure the flare energies.

An advantage of the GBTDS observing strategy compared to previous high-cadence photometric surveys such as TESS and Kepler, is the shorter effective exposure time. Although the photometric precision is lower, high-frequency signals are not averaged out by the photon noise added in the longer exposure. As a result, *short duration flares are detectable, even though they are not resolvable.* Although a single exposure is insufficient to derive a robust flare energy, the amplitude of the event will be sufficient for deriving a lower limit on the flare energy. A large sample of such lower limits when analyzed as a population will place strict constraints on higher energy flare frequencies. The fraction of detectable short duration flares can be approximated by the observing duty cycle, $t_{exp}$/cadence, which for the nominal GBTDS design is ~6%. Thus, in the context of this proposed statistical study, the observing cadence directly correlates to the detection efficiency of short duration flares, and a shorter cadence is always better for flare science. However, the read-out strategy section below describes a method for extracting short-timescale information from Roman without altering the survey cadence.

***Read-out strategy:*** Since the Roman Wide Field Instrument (WFI) H4RG-10 detectors can be read out non-destructively, i.e., without resetting the accumulated signal, the default read-out strategy will be to "sample up the ramp". This strategy records the accumulated voltage of each pixel every 3.04 seconds (referred to as a *read frame*) during an exposure, and then fits a slope to the read frames to infer the flux during the exposure (hereafter referred to as the "ramp-fitted flux"). This approach supplies several benefits, such as reduced read noise, efficient cosmic ray removal, and improved dynamic range (as the flux from stars that would otherwise saturate in a full exposure can be salvaged from earlier read frames). To limit data overhead, among other benefits, several consecutive read frames will be averaged onboard the spacecraft into a *resultant frame*, with the full exposure consisting of at least six resultant frames (Rauscher et al. 2019), from which the ramp-fitted flux will be inferred. The exact prescription determining which read frames will be averaged together for a resultant frame has yet to be decided, though the current candidate strategy will be to adopt unevenly sampled resultant frames (Casertano 2022, see also Figure 1).

Although Roman's cadence is not sensitive to resolving short duration flares, approximately 1 of every 15 flares (i.e., the duty cycle, see above) can still be detected. In these cases, performing photometry on resultant frames can provide additional constraints on the rise and fall of the flare, thereby breaking what would otherwise be a completely ambiguous degeneracy between the peak flux and event duration (or full-width at half maximum; FWHM), both of which are needed to infer the total event energy (see Figure 2). To test the ability of candidate readout strategies (i.e., the prescription for determining which read frames are averaged into resultant frames) to resolve flares we consider three scenarios: evenly sampled resultants (six resultant frames consisting each of three read frames), unevenly sampled resultants (six resultant frames each consisting of between 1-5 read frames), and the ideal case in which all read frames are preserved (gray points in all panels of Figure 2). As expected, a readout strategy that optimizes our ability to localize the peak flux of the flare is one with the highest number of resultant frames. For a set number of resultant frames, the optimal strategy is the unevenly sampled resultant frames, particularly if shorter resultant frames are at the beginning of the exposure. This strategy best constrains the shorter duration rise. Future studies will quantify this result further (Tovar Mendoza et al, *in prep*). However, regardless of the chosen GBTDS readout strategy, it is essential for flare science that all of the information from the resultant frames are made available.

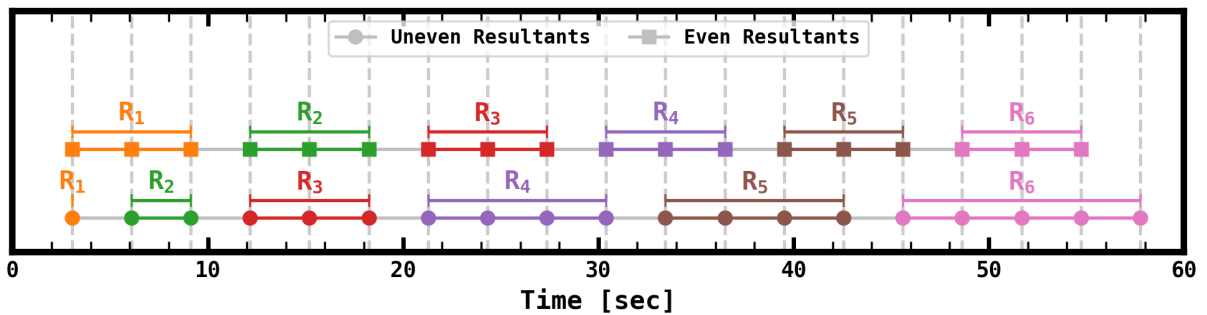

**Figure 1:** Candidate readout strategies introduced in Casertano (2022). Each individual point represents one read frame, and each color represents one resultant frame (e.g., $R_1$, $R_2$, etc.). Each resultant frame will consist of the average of one or more read frames.

*Filters:* A unique aspect of studying stellar flares with Roman is the wavelength regime. A high-precision, wide field time domain survey has never been conducted in the NIR. For stellar flares, this will enable us to probe a different part of the flare spectral energy distribution (e.g., Howard et al. 2020a; Maas et al. 2022). Flare effective temperatures are an active area of research (e.g., Howard et al. 2019) and the use of multiple filters would yield new information about the flare emission mechanisms at play (e.g., Osten et al. 2005; Paudel et al. 2021). Roman's wide filter F146 would enhance the overall signal-to-noise, however, the timescales of flares will be too short for Roman to observe a single flare in multiple filters. We expect bluer filters to have higher flare-star contrast ratios and improve the detectability of flares overall. However, redder filters are likely to improve our sensitivity with regard to lower-mass stars, reduce the effects of extinction, and probe deeper into the Galactic Bulge, exposing new and exciting stellar populations to flare demographics studies. The redder filters will also directly provide information on stellar flare contamination in the NIR (e.g., Davenport 2017), informing transiting exoplanet studies with JWST and future NASA missions. Even though Roman likely could not capture individual flares with multiple filters, observing similar populations of stars in multiple filters would enable a population-level study of flare emission mechanisms (e.g., Brasseur et al. 2022; Jackman et al. 2023).

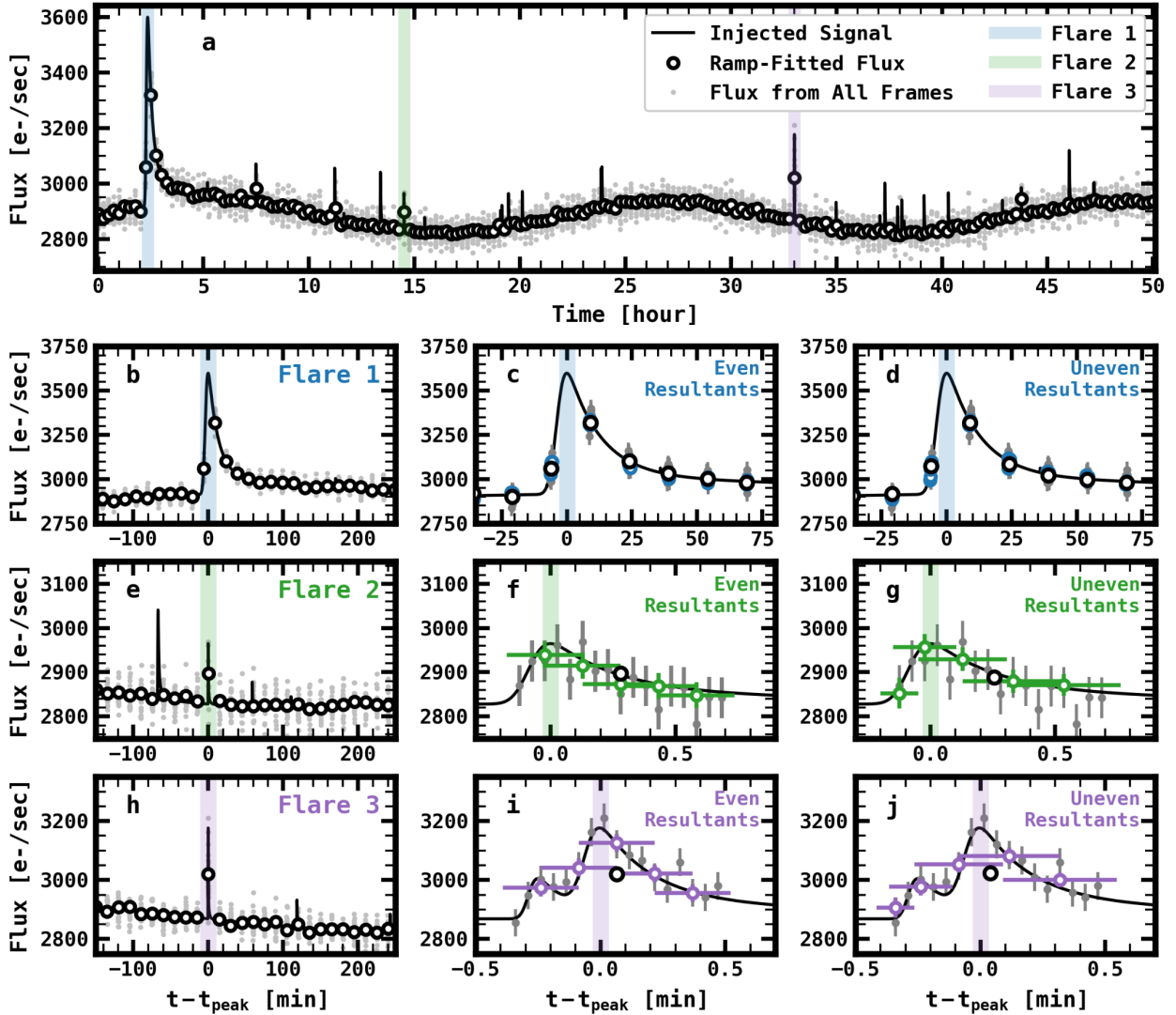

**Figure 2:** *Photometry from individual resultant frames can resolve high-frequency morphologies from large-amplitude flares.* **(a)** A simulated GBTDS light curve of a F146=19 mag$_{AB}$ star with rotational modulations of 2% in amplitude and dozens of flaring events of primarily short (<2 minutes) durations and large (0.1-10%) amplitudes (solid line), with three particular flares highlighted. The ramp-fitted flux (black points) is shown amongst the flux derived assuming all of the read frames can be preserved (gray points). Due to the 15-minute cadence, several high-amplitude, short-duration flares are missed. **(b-d)** A zoom-in on Flare 1, with the flux from six evenly sampled resultant frames (c) and six unevenly-sampled resultant frames (d) shown with colored points. The scenario in which all read frames are available are shown by the gray points. **(e-g)** The same as (b-d) but for Flare 2, where the flare FWHM is only 20 seconds. The colored error bars in the x-axis show the temporal extent of the resultant frames used to constrain the average flux. In this case, the unevenly sampled readout scheme (f) can resolve the rise of the flare, allowing us to localize the flare's peak thereby constraining its energy. **(h-j)** Same as (e-g) but for a complex, two-peaked flare. While detailed morphologies are still difficult

to constrain, the slow flare rise and wider FWHM hints at a complicated morphology that is better resolved in the unevenly-sampled readout scheme, and well-resolved if all read frames are available for analysis (gray points).